\documentclass[aip,rsi,reprint]{revtex4-1} 

\newcommand{\comment}[1]{}

\usepackage{amsmath}
\usepackage{amssymb}
\usepackage{latexsym}
\usepackage{indentfirst}
\usepackage{graphicx}
\usepackage{color}
\usepackage{bm}
\usepackage{multirow}

\begin{document}

\title{FLUXCAP: A flux-coupled ac/dc magnetizing device}

\author{D.~B. Gopman}
\email{daniel.gopman@physics.nyu.edu}
\author{H. Liu}
\author{A.~D. Kent}
\affiliation{Department of Physics, New York University, 4 Washington Place, New
             York, NY 10003, USA}

\begin{abstract}
We report on an instrument for applying ac and dc magnetic fields by
capturing the flux from a rotating permanent magnet and projecting it between two adjustable pole
pieces. This can be an alternative to standard electromagnets for experiments with small samples
or in probe stations in which an applied magnetic field is needed locally, 
with advantages that include a compact form-factor, very low power requirements and dissipation
as well as fast field sweep rates. This flux capture instrument (FLUXCAP) can
produce fields from $-400$ to $+400$~mT, with field resolution less than $1$~mT.
It generates static magnetic fields as well as ramped fields, with ramping rates
as high as $10$~T/s. We demonstrate the use of this apparatus for studying the magnetotransport 
properties of spin-valve nanopillars, a nanoscale device that exhibits giant magnetoresistance.
\end{abstract}

\pacs{07.55.-w,75.60.-d,75.47.De}

\maketitle

The effective synthesis and control of magnetic fields is of longstanding
fundamental interest for probing magnetic-field dependent phenomena. The ability
to effectively magnetize materials is also of tremendous technological importance
for testing magnetic devices, such as sensors, magnetic memories and other small magnetic elements, as
well as for characterizing a new generation of hybrid devices with semiconducting
and magnetic properties \cite{RefWorks:28,RefWorks:29,RefWorks:30}. This demand
for magnetizing devices has motivated the recent emergence of many different
methods for generating and directing magnetic fields
\cite{RefWorks:11,RefWorks:12,RefWorks:10,RefWorks:8,RefWorks:7,RefWorks:9}. From
static systems involving permanent magnets to electromagnetic systems built upon
current carrying coils and superconducting
magnets, there are many options when designing a source of magnetic fields. Most
designs consider: the maximum desired field, the field homogeneity in the sample
area, the size and access to the field region for probes (e.g. optical or electrical), the field sweep-rate 
and the magnet's linearity. In practice, the design factors are highly dependent on the researcher's aim.

Standard electromagnets have several disadvantages. The current-carrying electromagnets
are typically large, heavy units that can limit the optical access
to a device between the poles, consume high power in order to drive sufficient current through the
coils and require water cooling in order to
mitigate the Ohmic heating. 

In this paper, we introduce an ac/dc magnetizing device 
based on the coupling of magnetic flux into two parallel steel bars from a
diametrically magnetized permanent magnet that is mounted to a rotating stage. We
denote this instrument FLUXCAP for its flux capture characteristics.
The FLUXCAP is an alternative to standard magnetizing devices for generating the
fields between two soft pole pieces. It has been designed for testing small magnetic devices,
whose lateral size is much smaller than the diameter of the pole pieces. This apparatus has several
advantages over standard magnets: it is portable and can
run entirely on battery power; the only Ohmic losses are in the motor,
for which overheating can easily be prevented by heat-sinking; and the
high speed by which the magnet can be rotated permits higher field ramping rates
than many electromagnets which typically have a large inductance. Furthermore, the FLUXCAP permits 
optical access and is high vacuum compatible, permitting a wide range of test applications. 

\begin{figure}[t!]
  \begin{center}
    \includegraphics[width=3.375in,
    keepaspectratio=True]
    {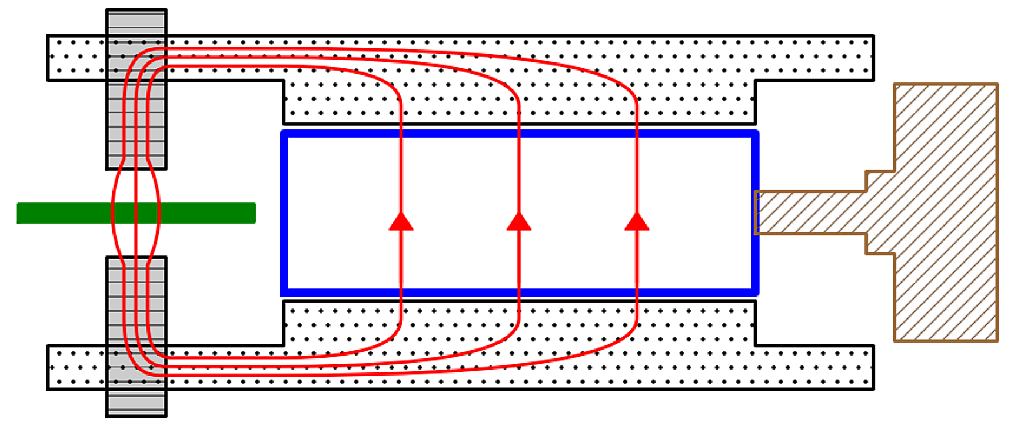}
  \end{center}
  \caption{\label{fig:diagram} Diagram of a flux-coupled dipolar magnet based on
  a permanent magnet rod with diametric magnetization mounted to a motor. The
  permanent magnet is indicated by the blue rectangle. The motor is indicated
  by the brown hashed region, and the steel bars by
  the black dotted regions. The pole pieces (solid grey with black stripes) are adjustable for changing the gap length. 
  A test device is drawn between the two pole pieces
  as the green solid rectangle. The red closed curves symbolize the
  magnetic field lines in the system.}
\end{figure}

\section{Physical Description and Principles of Operation}

Figure~\ref{fig:diagram} illustrates our implementation of a FLUXCAP magnet. The
yoke of the FLUXCAP is a Neodymium Iron Boron (NIB) magnetic rod (diameter 0.5
inches, length 2 inches) \cite{NIB} as indicated by the blue rectangle in the figure. It is
magnetized uniformly along the rod diameter. The NIB
magnet is attached to a motor (brown hashed) which permits continuous rotation of the rod about
its long axis, and consequently, continuous rotation of the magnetization in the
plane perpendicular to the rotation axis. The magnet yoke is flanked on both sides by
soft pole pieces -- two low-carbon steel bars as indicated by the black dotted
regions (0.25~$\mathrm{in}^2$ square cross-section, length 6 inches). Two
threaded holes near the termination of the bars accommodate one-quarter inch
threaded steel rods, completing the pole pieces. The pole gap is adjusted by
threading the removable pole pieces (grey with black stripes) into and out of the threaded holes in the steel bars. Test devices, as indicated by
the green solid rectangle, are inserted in the gap between the pole pieces and a
commercial Gaussmeter is placed at the sample location
to monitor the field produced between the poles. This entire apparatus is lightweight, weighing less 
than 10~kg.

The soft pole pieces capture the flux incident from the yoke and focus the
field lines across the relatively short air gap between the pole pieces.
As the NIB rod is rotated on its axis, the net flux captured into the pole pieces
from the yoke varies periodically. This rotation translates into a
nearly sinusoidally varying magnetic field between the two poles.

The operation of the FLUXCAP magnet depends upon the capture of magnetic flux
from a permanent magnet into two parallel steel bars placed on each side of the
magnet. Maximal flux transfer occurs when the magnetization of the diametrically
magnetized permanent magnet is directed toward the faces of the steel bars and
minimal flux is transferred when the magnetization is oriented perpendicular to
the faces of the bars. Thus, the permanent magnet is rotated by a motor
in order to vary the flux captured by the steel bars by varying the angle between
the magnetization direction and the steel bar faces.

\begin{figure}[b]
  \begin{center}
    \includegraphics[width=3.375in,
    keepaspectratio=True]
    {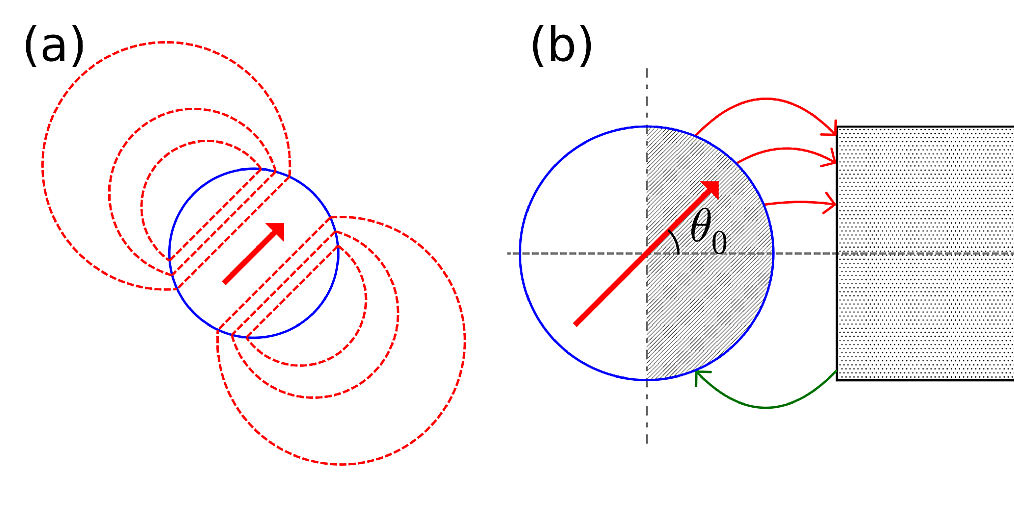}
  \end{center}
  \caption{\label{fig:netFLUX} Distortion of the magnetic field lines from a
  diametrically magnetized cylinder due to the proximity of a steel bar. The
  red arrows in both (a) and (b) represent the radial magnetization direction.
  The red dashed curves in (a) represent the magnetic field lines. The angle
  $\theta_0$ in (b) is the angle between magnetization and the steel surface
  normal. The red and green curves represent the magnetic field lines coming
  out from and into the magnetic rod.}
\end{figure}

We present a model for understanding the basic dependence of the flux in the
steel rods as a function of the magnetization direction of the permanent magnet.
Equation~\eqref{radialField} presents the field from an infinite uniformly
magnetized rod with diametric magnetization in cylindrical coordinates:
\begin{equation}
\textbf{B} = \left\{\begin{array}{rl}
 B_{\mathrm max}( \widehat{\rho} \cos \phi -
\widehat{\phi} \sin \phi), &\ \rho < R \\  B_{\mathrm max}\left(\frac{\displaystyle
R}{\displaystyle \rho}\right)^2 (\widehat{\rho} \cos \phi + \widehat{\phi} \sin
\phi), &\ \rho > R. \end{array}\right.
\label{radialField}
\end{equation}
Here $B_{\mathrm max} = \mu_0M_s/2$,
where $M_s$ is the saturation magnetization of the permanent magnet. $\rho$ and
$\phi$ are the radial and angular cylindrical coordinates. $R$ is
the radius of the permanent magnet rod. The geometry of this arrangement is
further depicted in Fig.~\ref{fig:netFLUX}(a). While the real magnet is
finite in extent, we believe that this approximately captures the relevant
behavior of this system because the length of the magnetic rod is much larger
than the distance between the rod and the steel bars (approximately one-eighth of
an inch). We are ultimately interested in the field lines extending radially
outward and into the steel piece, which is set by 
$B_{\mathrm max}$, the maximum magnetic field at the surface of the magnet. This value can be
directly measured with a magnetic field sensor placed on the surface of the
magnet, which we have measured as 0.993~T. Having established an expression for the field, we proceed to a
description of the flux in the steel bars.

The proximity of the two parallel steel bars have a non-negligible effect on the
fields from the permanent magnet. We assume the magnetic field on the surface of
the permanent magnet is left unchanged, but that the magnetic field lines are
distorted in such a way that field lines on the right semicircular face of the
magnet terminate on the right steel bar and lines on the left semicircular face
terminate on the left steel bar, as depicted graphically in
Fig.~\ref{fig:netFLUX}(b). Therefore, we can estimate the total magnetic flux
into the right steel bar as the net magnetic flux exiting the right semicircular
face of the magnet. Equation~\eqref{fluxTOT} gives the total flux $\Phi$ as a function
of $\theta_0$, the angle between the magnetization and the steel surface normal:
\begin{equation}
\Phi = B_{max} \cdot 2R \ell \cdot \cos \theta_0,
\label{fluxTOT}
\end{equation}
where $\ell$ is the length of the magnet.

Finally, the magnetic flux is directed toward two pole pieces extending into the
gap between the parallel steel bars. For sufficient distances between the
permanent magnet and the pole pieces, all of the field between the pole pieces is indirectly coupled
through the flux in the steel bars. The magnitude of this field is inversely
proportional to the surface area of the pole tips. It is also sensitive to the
pole gap and distance of the pole pieces from the magnet, both of which can
contribute to flux losses through leakage along the gap between the parallel bars and fringing at the poles. We present
an implementation of this method of flux capture and direction that exploits a
pole displacement and tip surface area that gives a fraction of a Tesla magnetic field
in a $\sim 1$ cm$^3$ volume. We also adjust the pole gap in order to control the peak field applied between
the poles, as discussed below.

\section{FLUXCAP Operation}

\subsection{Variable Amplitude and Precision Control}

The gap between two steel pole pieces can be adjusted to change the maximum applied field. 
This varies the peak field amplitude of the alternating magnetic field when the magnet is rotated continuously. 
Reducing the peak field amplitude may be useful in studying devices that have multiple magnetic layers, 
some of which are not intended to be remagnetized.

Reducing the peak field amplitude may also be useful in applications where field
precision is of most importance. For example, the FLUXCAP
can be used to generate dc magnetic fields with the magnet positioned using a
stepper motor. The field precision is related to the minimum rotation that the
motor can produce and the maximum field amplitude. Each finite step from a stepper
motor (typically 0.9 degrees) corresponds to a change in field of just over one
percent of the peak amplitude. Therefore, larger pole separations may be
desirable to obtain finer control over magnetic fields.

\begin{figure}[t]
  \begin{center}
    \includegraphics[width=3.375in,
    keepaspectratio=True]
    {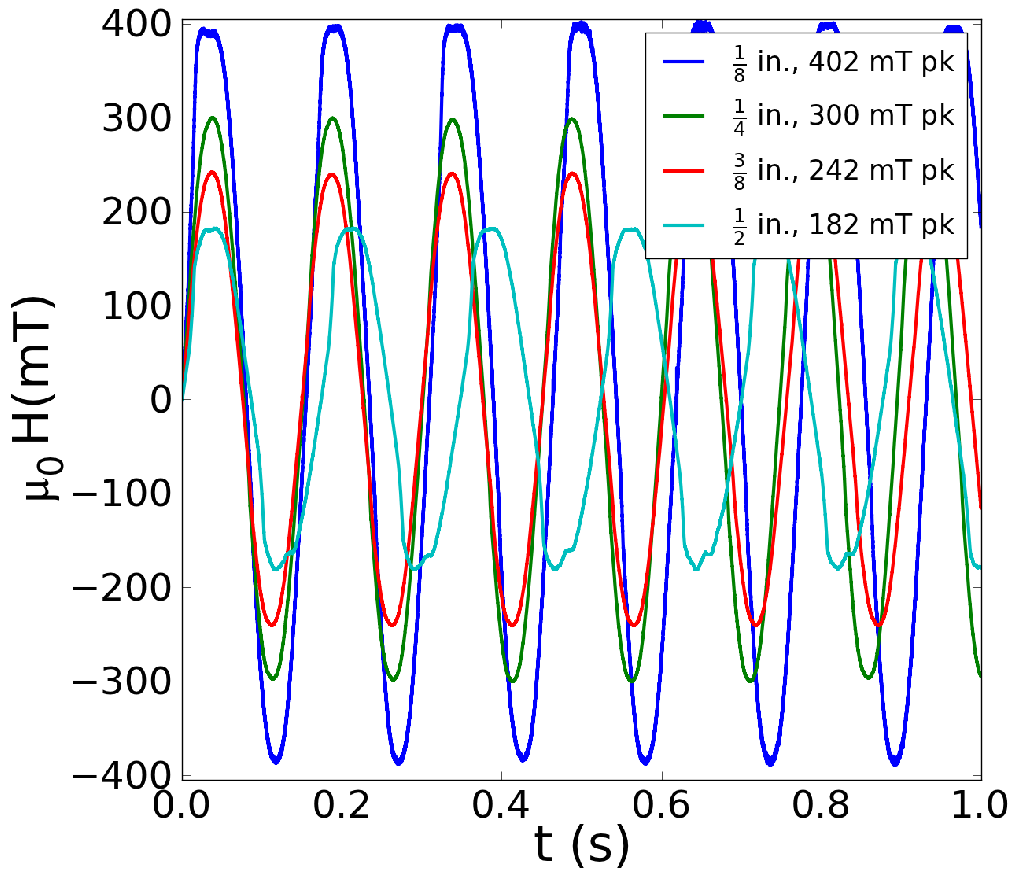}
  \end{center}
  \caption{\label{fig:PeakAmplitude} Magnetic field between the pole pieces
  as a function of time as the permanent magnet is rotated continuously at a rate of
  400~rpm. The maximum applied field is seen to be a function of the gap spacing.}
\end{figure}

Figure~\ref{fig:PeakAmplitude} demonstrates the ability to adjust the maximum
field between the poles. We adjusted the 0.25 inch diameter threaded rods to create pole gaps in
0.125 inch increments between 0.125 and 0.5 inches and operated the FLUXCAP using
a 12~V battery-powered dc motor. As the permanent magnet was rotated continuously at a rate of 400~revolutions per minute, 
field measurements were made using a commercial
Gaussmeter and then digitized at 48~kHz. We demonstrate control over the peak
field amplitude from 400~mT down to 180~mT. 

The peak amplitude $B_{pk}$ at the pole gap of 0.125 inches allows us to estimate the efficiency by which
we capture the flux from our NIB yoke into the two pole pieces. Assuming an effective fringing area $A_f$ about 50\% larger
than the pole face, we compute the efficiency, $e = B_{pk}A_f / \Phi$, to be approximately 15\%. We estimate that 85\% of the 
flux is being lost to leakage across the space between the steel bars. The FLUXCAP could be made more efficient by employing higher permeability materials and through optimizing the pole piece geometry.

\subsection{Variable frequency and Field Ramping}

\begin{figure}[b]
  \begin{center}
    \includegraphics[width=3.375in,
    keepaspectratio=True]
    {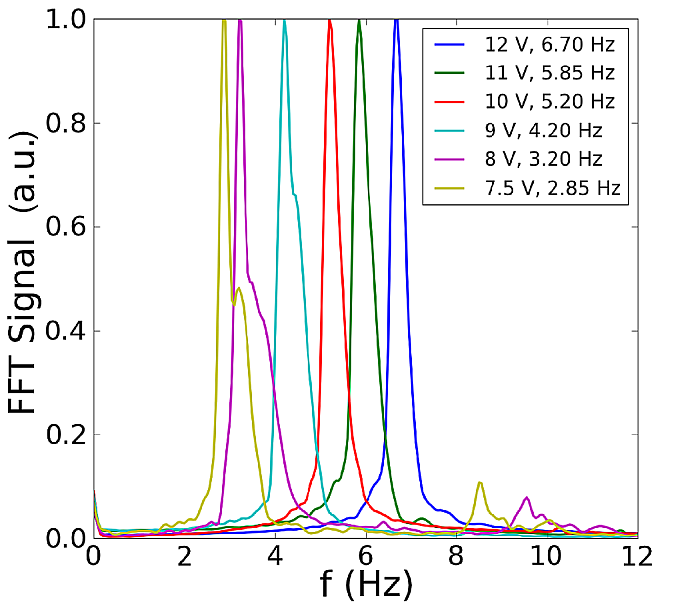}
  \end{center}
  \caption{\label{fig:SPECTRA} FFT spectra indicating the various magnet
  rotation rates.}
\end{figure}

\begin{figure*}
    \includegraphics[width=6.69in,
    keepaspectratio=True]
    {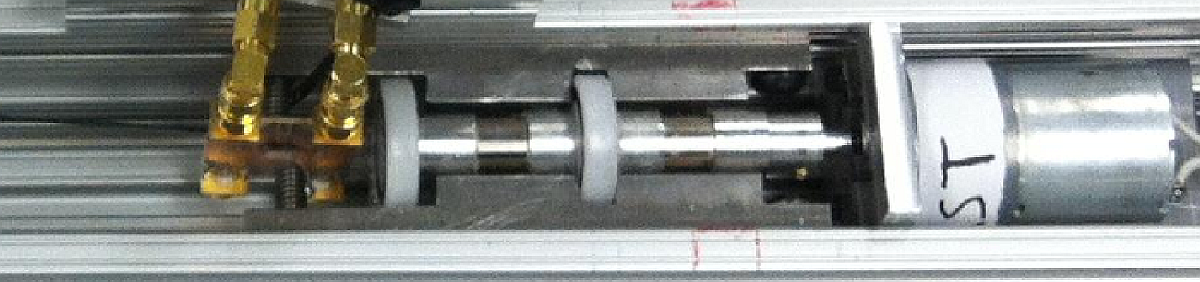}
  \caption{\label{fig:Magnet} FLUXCAP apparatus in a testing configuration. The FLUXCAP motor and steel bars are bolted to aluminum tracks. The NIB yoke is encased in aluminum shells for coupling to the motor with a brass set screw and to two plastic ball bearings at the midpoint and the endpoint of the magnet. Adjustable pole pieces emerge from the steel bars where the test sample has been clamped to the aluminum track and a gaussmeter probe is attached behind the sample.
  and sample.
  }
\end{figure*}

The FLUXCAP magnet can operate either as a stationary or
an alternating field magnetizing device. In the alternating field operation mode,
a dc motor generates continuous rotation of the NIB yoke which drives an
alternating magnetic field between the poles. By changing the frequency of
rotation, a variation of the field sweeping rate (frequency) can be achieved.
Figure~\ref{fig:SPECTRA} shows the Fast Fourier Transform spectra of the
alternating magnetic field between the poles of the FLUXCAP under different
rotation speeds of the motor. Measurements were taken under a 0.25 inch pole gap
using the same field acquisition methods described above. We demonstrate
alternating magnetic fields with frequencies ranging between 3~Hz and 7~Hz for
voltages ranging from 7~V to 12~V placed across the motor. 

For the lower frequencies, the FFT spectra show wide sidebands due to a varying rotation rate of the permanent magnet. We used a 12~V/84~oz-in 37~mm dc motor~\cite{Pololu.DCMotor}, which uses 12~W of power (1~A or 20\% of its stall current) at 12~V. At lower voltages, the maximal output torque of the motor decreases and rotating the magnet away from the steel bars requires larger torque. For inputs below 7~V, this motor stalls. We use a higher torque stepper motor when slower field ramp rates are needed~\cite{Lin.Stepper}.

The frequency of the alternating magnetic field corresponds to an effective
linear ramping rate over $\pm$ 85 percent of the maximum amplitude field. For the
frequencies given here and the 0.3~T peak amplitude for a 0.25~inch pole
separation, the ramping rates vary from 10~T/s
down to 4~T/s for the lowest rotation frequency. These high field ramping rates
make the FLUXCAP an efficient rapid magnetizing device when used with a 
continuously rotating motor. Slower variation of the field has been achieved with the use of
a high torque stepper motor, permitting field ramping rates to decreasing by
several orders of magnitude. This could be relevant to studies of
thermally-activated magnetization reversal in which a magnet's coercivity is (typically
logarithmically)
sensitive to the sweeping rate
\cite{RefWorks:19,RefWorks:14,RefWorks:15,RefWorks:17,RefWorks:16,RefWorks:13}.

\section{Application: Fast magnetizing of spin-valve nanopillars}

The FLUXCAP magnet can facilitate fast characterization of many magnetic devices
such as spin-valve nanopillars -- a two terminal magnetic device composed of two
ferromagnetic layers separated by a thin non-magnetic layer \cite{RefWorks:20,RefWorks:14}. Typically a
spin-valve device exhibits two stable resistance states depending on the relative
magnetization orientation of the two magnetic layers from the Giant
Magnetoresistance (GMR) effect \cite{RefWorks:22,RefWorks:23,RefWorks:24,RefWorks:25}. The spin-valve state can
be toggled between high (antiparallel) and low (parallel) resistance by applied
magnetic fields. Characteristic of spin-valve nanopillars is the use of
ferromagnetic layers with different coercivities, such that one ferromagnet is typically
fixed (a reference layer) while the other ferromagnet (the free layer) can be switched
relative to the reference magnet. We can determine the relative orientation of
the layers by measuring the device resistance as a function of the applied
magnetic field. More critically, using the 7~Hz rotation rate of the FLUXCAP, we
can rapidly conduct MR hysteresis loops to measure the coercivity of the free
layer and the giant magnetoresistance of the spin-valve. The FLUXCAP also could be
incorporated into a probe setup for characterizing the properties of spin-valve and
magnetic tunnel juction (MTJ) devices.

\begin{figure}[b!]
  \begin{center}
    \includegraphics[width=3.375in,
    keepaspectratio=True]
    {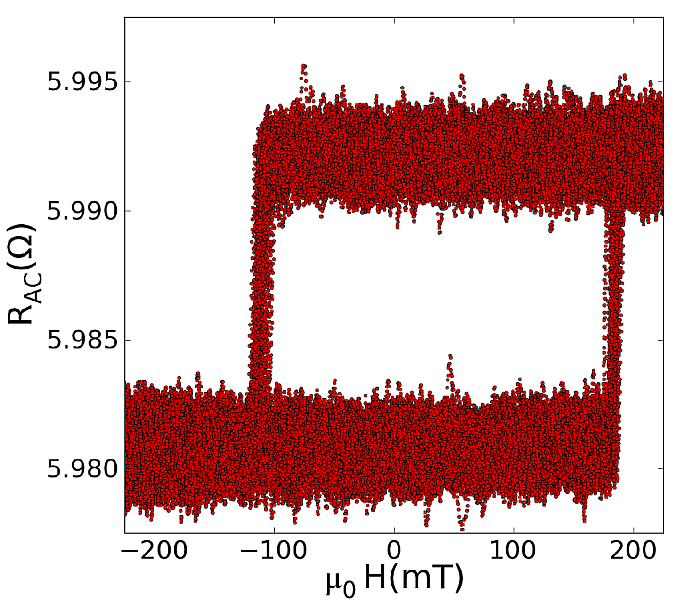}
  \end{center}
  \caption{\label{fig:130loops} GMR signal versus field for 130 hysteresis loops
  obtained in 20 seconds.}
\end{figure}

Figure~\ref{fig:Magnet} demonstrates a testing configuration for this apparatus.
A spin-valve nanopillar device is wire bonded to a coplanar waveguide board,
which is in turn has been soldered to end-launch coaxial jacks. The waveguide is
mounted rigidly to the outer Aluminum rail of the apparatus such that the device
is centered between the two pole pieces. A commercial Gaussmeter probe is also
mounted on the aluminum rail and is attached to one of the pole pieces. A small
ac excitation current probes the differential resistance across the 300 $\times$
50~$\mathrm{nm}^2$ spin-valve nanopillar, whose physical properties have
been described in detail elsewhere~\cite{RefWorks:20}. The FLUXCAP is
configured with a one-quarter inch pole gap and 12~V power to run the motor at 7~Hz.

\begin{figure}[b!]
  \begin{center}
    \includegraphics[width=3.375in,
    keepaspectratio=True]
    {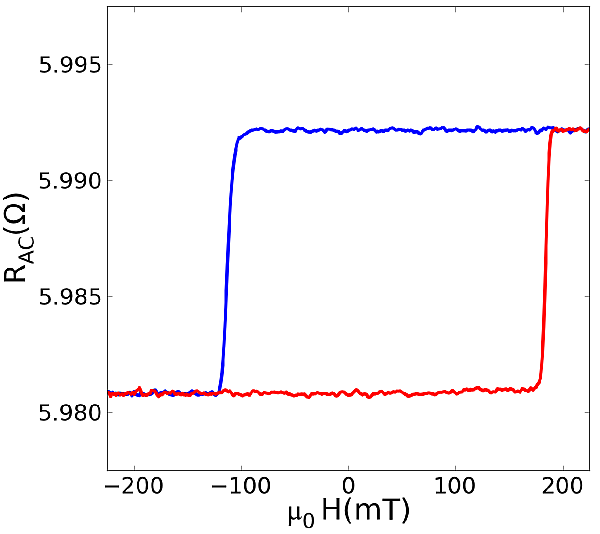}
  \end{center}
  \caption{\label{fig:HYST.avg} Averaged magnetic hysteresis loop of a spin-valve
  device.}
\end{figure}

Figure~\ref{fig:130loops} shows over 100 hysteresis loops recorded from 20
seconds of operating the FLUXCAP. Sharp changes in the differential resistance
$R_{AC}$ indicate toggling of the magnetization of the free layer from ``up''
(anti-parallel) to ``down'' (parallel) relative to the reference layer. As
mentioned previously, the reference layer has a coercivity over 1~T, and is kept
fixed during these measurements. Due to the thermally activated nature of
magnetization reversal, a characteristic distribution of switching fields is
apparent in this ensemble of hysteresis loops. It is therefore effective to
consider an averaged hysteresis loop, such as the one depicted in
Fig~\ref{fig:HYST.avg}. From this figure, we estimate a coercivity of 150~mT and
a GMR ratio ($\Delta R/ R$) of 0.2\%, which is consistent
with similar devices \cite{RefWorks:14,RefWorks:26,RefWorks:27}.

\section{Conclusions}

We have demonstrated the operation of the FLUXCAP, a compact magnetizing device
based upon the capture and focusing of flux from a rotating permanent 
magnet. This device can perform most of the same tasks as conventional
electromagnet based magnetizing devices in that it can synthesize static
and dynamic magnetic fields over a broad range of field values. Yet the FLUXCAP
immediately presents itself as an elegant alternative: it operates as an 
ac magnetizing device requiring only a 12~V battery and a dc motor; its power consumption is marginal (12~W),
and it does not require water cooling. The pole pieces are modular - it is straightforward to change the maximum 
field applied by varying the pole gap or even substituting a threaded rod with a different bevel or chamfer. This permits easy modifications to the magnitude and homogeneity of the applied field with minor changes in the FLUXCAP design.
Furthermore, large field ramp rates are possible with FLUXCAP and
have been demonstrated for studying spin-valve nanopillar devices. This enables
statistical studies of thermally activated magnetization reversal, quick
resetting of magnetic devices and testing the dynamic response of magnetic field
sensors.

FLUXCAP magnets are versatile enough to function well in a variety of other
applications. The setup could be made UHV compatible
 -- in fact, the permanent magnet and
motor could be placed outside of the UHV chamber and the flux coupled into the
chamber with the soft steel core. FLUXCAP magnets can clearly be used
for electronic transport measurements and could be integrated into probe stations. For example,
it could be used to add magnetic capabilities to a semiconductor tester.
By designing the shape of pole pieces and the position to place the
sample, one can achieve different field directions using the
FLUXCAP. Finally, it is easy to imagine combining two or three such magnets to
generate a two-dimensional or even three-dimensional vector field for
sophisticated measurements.

\section*{Acknowledgments}
We appreciate Dr. St\'{e}phane Mangin of Nancy Universit\'{e} and Dr. Eric E. Fullerton of the University of California, San Diego for providing the spin-valve samples used in the characterization of magnetic properties used in this study. We would also like to acknowledge Dr. James Rantschler of Xavier University of Louisiana for fruitful discussions leading to the design of FLUXCAP. This research was supported by NSF Grant No. DMR-1006575.


\begin{thebibliography}{25}
\expandafter\ifx\csname natexlab\endcsname\relax\def\natexlab#1{#1}\fi
\expandafter\ifx\csname bibnamefont\endcsname\relax
  \def\bibnamefont#1{#1}\fi
\expandafter\ifx\csname bibfnamefont\endcsname\relax
  \def\bibfnamefont#1{#1}\fi
\expandafter\ifx\csname citenamefont\endcsname\relax
  \def\citenamefont#1{#1}\fi
\expandafter\ifx\csname url\endcsname\relax
  \def\url#1{\texttt{#1}}\fi
\expandafter\ifx\csname urlprefix\endcsname\relax\def\urlprefix{URL }\fi
\providecommand{\bibinfo}[2]{#2}
\providecommand{\eprint}[2][]{\url{#2}}

\bibitem[{\citenamefont{Hueso and Dediu}(2009)}]{RefWorks:28}
\bibinfo{author}{\bibfnamefont{L.}~\bibnamefont{Hueso}} \bibnamefont{and}
  \bibinfo{author}{\bibfnamefont{V.~A.} \bibnamefont{Dediu}},
  \bibinfo{journal}{Nature Materials} \textbf{\bibinfo{volume}{8}},
  \bibinfo{pages}{707} (\bibinfo{year}{2009}).

\bibitem[{\citenamefont{Yi et~al.}(2010)\citenamefont{Yi, Lim, Xing, Fan, Van,
  Huang, Yang, Huang, Qin, Wang et~al.}}]{RefWorks:29}
\bibinfo{author}{\bibfnamefont{J.~B.} \bibnamefont{Yi}},
  \bibinfo{author}{\bibfnamefont{C.~C.} \bibnamefont{Lim}},
  \bibinfo{author}{\bibfnamefont{G.~Z.} \bibnamefont{Xing}},
  \bibinfo{author}{\bibfnamefont{H.~M.} \bibnamefont{Fan}},
  \bibinfo{author}{\bibfnamefont{L.~H.} \bibnamefont{Van}},
  \bibinfo{author}{\bibfnamefont{S.~L.} \bibnamefont{Huang}},
  \bibinfo{author}{\bibfnamefont{K.~S.} \bibnamefont{Yang}},
  \bibinfo{author}{\bibfnamefont{X.~L.} \bibnamefont{Huang}},
  \bibinfo{author}{\bibfnamefont{X.~B.} \bibnamefont{Qin}},
  \bibinfo{author}{\bibfnamefont{B.~Y.} \bibnamefont{Wang}},
  \bibnamefont{et~al.}, \bibinfo{journal}{Physical Review Letters}
  \textbf{\bibinfo{volume}{104}}, \bibinfo{pages}{137201}
  (\bibinfo{year}{2010}).

\bibitem[{\citenamefont{Wang et~al.}(2012)\citenamefont{Wang, Macia,
  Wohlgenannt, Kent, and Flatte}}]{RefWorks:30}
\bibinfo{author}{\bibfnamefont{F.}~\bibnamefont{Wang}},
  \bibinfo{author}{\bibfnamefont{F.}~\bibnamefont{Macia}},
  \bibinfo{author}{\bibfnamefont{M.}~\bibnamefont{Wohlgenannt}},
  \bibinfo{author}{\bibfnamefont{A.~D.} \bibnamefont{Kent}}, \bibnamefont{and}
  \bibinfo{author}{\bibfnamefont{M.~E.} \bibnamefont{Flatte}},
  \bibinfo{journal}{Physical Review X} \textbf{\bibinfo{volume}{2}},
  \bibinfo{pages}{021013} (\bibinfo{year}{2012}).

\bibitem[{\citenamefont{Chang et~al.}(2010)\citenamefont{Chang, Chen, and
  Hwang}}]{RefWorks:11}
\bibinfo{author}{\bibfnamefont{W.-H.} \bibnamefont{Chang}},
  \bibinfo{author}{\bibfnamefont{J.-H.} \bibnamefont{Chen}}, \bibnamefont{and}
  \bibinfo{author}{\bibfnamefont{L.-P.} \bibnamefont{Hwang}},
  \bibinfo{journal}{Magnetic Resonance Imaging} \textbf{\bibinfo{volume}{28}}
  (\bibinfo{year}{2010}).

\bibitem[{\citenamefont{Arena et~al.}(2009)\citenamefont{Arena, Ding, Vescovo,
  Zohar, Guan, and Bailey}}]{RefWorks:12}
\bibinfo{author}{\bibfnamefont{D.~A.} \bibnamefont{Arena}},
  \bibinfo{author}{\bibfnamefont{Y.}~\bibnamefont{Ding}},
  \bibinfo{author}{\bibfnamefont{E.}~\bibnamefont{Vescovo}},
  \bibinfo{author}{\bibfnamefont{S.}~\bibnamefont{Zohar}},
  \bibinfo{author}{\bibfnamefont{Y.}~\bibnamefont{Guan}}, \bibnamefont{and}
  \bibinfo{author}{\bibfnamefont{W.~E.} \bibnamefont{Bailey}},
  \bibinfo{journal}{Review of Scientific Instruments}
  \textbf{\bibinfo{volume}{80}}, \bibinfo{pages}{083903}
  (\bibinfo{year}{2009}).

\bibitem[{\citenamefont{Gilbert et~al.}(2012)\citenamefont{Gilbert, Mertins,
  Tesch, Berges, Feilbach, and Schneider}}]{RefWorks:10}
\bibinfo{author}{\bibfnamefont{M.}~\bibnamefont{Gilbert}},
  \bibinfo{author}{\bibfnamefont{H.~C.} \bibnamefont{Mertins}},
  \bibinfo{author}{\bibfnamefont{M.}~\bibnamefont{Tesch}},
  \bibinfo{author}{\bibfnamefont{O.}~\bibnamefont{Berges}},
  \bibinfo{author}{\bibfnamefont{H.}~\bibnamefont{Feilbach}}, \bibnamefont{and}
  \bibinfo{author}{\bibfnamefont{C.~M.} \bibnamefont{Schneider}},
  \bibinfo{journal}{Review of Scientific Instruments}
  \textbf{\bibinfo{volume}{83}}, \bibinfo{pages}{025109}
  (\bibinfo{year}{2012}).

\bibitem[{\citenamefont{Heigl et~al.}(2002)\citenamefont{Heigl, Krupin, Kaindl,
  and Starke}}]{RefWorks:8}
\bibinfo{author}{\bibfnamefont{F.}~\bibnamefont{Heigl}},
  \bibinfo{author}{\bibfnamefont{O.}~\bibnamefont{Krupin}},
  \bibinfo{author}{\bibfnamefont{G.}~\bibnamefont{Kaindl}}, \bibnamefont{and}
  \bibinfo{author}{\bibfnamefont{K.}~\bibnamefont{Starke}},
  \bibinfo{journal}{Review of Scientific Instruments}
  \textbf{\bibinfo{volume}{73}}, \bibinfo{pages}{369} (\bibinfo{year}{2002}).

\bibitem[{\citenamefont{Li et~al.}(2012)\citenamefont{Li, Jin, Son, Tan, Cao,
  Hwang, and Qiu}}]{RefWorks:7}
\bibinfo{author}{\bibfnamefont{J.}~\bibnamefont{Li}},
  \bibinfo{author}{\bibfnamefont{E.}~\bibnamefont{Jin}},
  \bibinfo{author}{\bibfnamefont{H.}~\bibnamefont{Son}},
  \bibinfo{author}{\bibfnamefont{A.}~\bibnamefont{Tan}},
  \bibinfo{author}{\bibfnamefont{W.~N.} \bibnamefont{Cao}},
  \bibinfo{author}{\bibfnamefont{C.}~\bibnamefont{Hwang}}, \bibnamefont{and}
  \bibinfo{author}{\bibfnamefont{Z.~Q.} \bibnamefont{Qiu}},
  \bibinfo{journal}{Review of Scientific Instruments}
  \textbf{\bibinfo{volume}{83}}, \bibinfo{pages}{033906}
  (\bibinfo{year}{2012}).

\bibitem[{\citenamefont{Nolle et~al.}(2012)\citenamefont{Nolle, Weigand,
  Audehm, Goering, Wiesemann, Wolter, Nolle, and Schuetz}}]{RefWorks:9}
\bibinfo{author}{\bibfnamefont{D.}~\bibnamefont{Nolle}},
  \bibinfo{author}{\bibfnamefont{M.}~\bibnamefont{Weigand}},
  \bibinfo{author}{\bibfnamefont{P.}~\bibnamefont{Audehm}},
  \bibinfo{author}{\bibfnamefont{E.}~\bibnamefont{Goering}},
  \bibinfo{author}{\bibfnamefont{U.}~\bibnamefont{Wiesemann}},
  \bibinfo{author}{\bibfnamefont{C.}~\bibnamefont{Wolter}},
  \bibinfo{author}{\bibfnamefont{E.}~\bibnamefont{Nolle}}, \bibnamefont{and}
  \bibinfo{author}{\bibfnamefont{G.}~\bibnamefont{Schuetz}},
  \bibinfo{journal}{Review of Scientific Instruments}
  \textbf{\bibinfo{volume}{83}}, \bibinfo{pages}{046112}
  (\bibinfo{year}{2012}).

\bibitem[{NIB()}]{NIB}
\bibinfo{note}{This study used Neodymium Iron Boron (NIB) magnets purchased
  from K \& J Magnetics, http://www.kjmagnetics.com}.

\bibitem[{Pololu.DCMotor()}]{Pololu.DCMotor}
\bibinfo{note}{This study used a 12~V 19:1 Metal DC Gearmotor (37mm Shaft) with 84 oz-in torque and 500 rpm. The motor is from Pololu Metal Gearmotors, http://www.pololu.com}.

\bibitem[{Lin.Stepper()}]{Lin.Stepper}
\bibinfo{note}{For slower field sweeping we use a High Torque (175 oz-in) Stepper Motor. It can be found under model number 5709M from Lin Engineering, http://www.linengineering.com}.

\bibitem[{\citenamefont{Wernsdorfer et~al.}(1997)\citenamefont{Wernsdorfer,
  Orozco, Hasselbach, Benoit, Barbara, Demoncy, Loiseau, Pascard, and
  Mailly}}]{RefWorks:19}
\bibinfo{author}{\bibfnamefont{W.}~\bibnamefont{Wernsdorfer}},
  \bibinfo{author}{\bibfnamefont{E.~B.} \bibnamefont{Orozco}},
  \bibinfo{author}{\bibfnamefont{K.}~\bibnamefont{Hasselbach}},
  \bibinfo{author}{\bibfnamefont{A.}~\bibnamefont{Benoit}},
  \bibinfo{author}{\bibfnamefont{B.}~\bibnamefont{Barbara}},
  \bibinfo{author}{\bibfnamefont{N.}~\bibnamefont{Demoncy}},
  \bibinfo{author}{\bibfnamefont{A.}~\bibnamefont{Loiseau}},
  \bibinfo{author}{\bibfnamefont{H.}~\bibnamefont{Pascard}}, \bibnamefont{and}
  \bibinfo{author}{\bibfnamefont{D.}~\bibnamefont{Mailly}},
  \bibinfo{journal}{Physical Review Letters} \textbf{\bibinfo{volume}{78}}
  (\bibinfo{year}{1997}).

\bibitem[{\citenamefont{Gopman et~al.}(2012)\citenamefont{Gopman, Bedau,
  Mangin, Lambert, Fullerton, Katine, and Kent}}]{RefWorks:14}
\bibinfo{author}{\bibfnamefont{D.~B.} \bibnamefont{Gopman}},
  \bibinfo{author}{\bibfnamefont{D.}~\bibnamefont{Bedau}},
  \bibinfo{author}{\bibfnamefont{S.}~\bibnamefont{Mangin}},
  \bibinfo{author}{\bibfnamefont{C.~H.} \bibnamefont{Lambert}},
  \bibinfo{author}{\bibfnamefont{E.~E.} \bibnamefont{Fullerton}},
  \bibinfo{author}{\bibfnamefont{J.~A.} \bibnamefont{Katine}},
  \bibnamefont{and} \bibinfo{author}{\bibfnamefont{A.~D.} \bibnamefont{Kent}},
  \bibinfo{journal}{Applied Physics Letters} \textbf{\bibinfo{volume}{100}},
  \bibinfo{pages}{062404} (\bibinfo{year}{2012}).

\bibitem[{\citenamefont{Jiang et~al.}(2003)\citenamefont{Jiang, Abe, Nozaki,
  Tezuka, and Inomata}}]{RefWorks:15}
\bibinfo{author}{\bibfnamefont{Y.}~\bibnamefont{Jiang}},
  \bibinfo{author}{\bibfnamefont{S.}~\bibnamefont{Abe}},
  \bibinfo{author}{\bibfnamefont{T.}~\bibnamefont{Nozaki}},
  \bibinfo{author}{\bibfnamefont{N.}~\bibnamefont{Tezuka}}, \bibnamefont{and}
  \bibinfo{author}{\bibfnamefont{K.}~\bibnamefont{Inomata}},
  \bibinfo{journal}{Physical Review B} \textbf{\bibinfo{volume}{68}},
  \bibinfo{pages}{224426} (\bibinfo{year}{2003}).

\bibitem[{\citenamefont{Sun et~al.}(2002{\natexlab{a}})\citenamefont{Sun, Chen,
  Suzuki, Parkin, and Koch}}]{RefWorks:17}
\bibinfo{author}{\bibfnamefont{J.~Z.} \bibnamefont{Sun}},
  \bibinfo{author}{\bibfnamefont{L.}~\bibnamefont{Chen}},
  \bibinfo{author}{\bibfnamefont{Y.}~\bibnamefont{Suzuki}},
  \bibinfo{author}{\bibfnamefont{S.~S.~P.} \bibnamefont{Parkin}},
  \bibnamefont{and} \bibinfo{author}{\bibfnamefont{R.~H.} \bibnamefont{Koch}},
  \bibinfo{journal}{Journal of Magnetism and Magnetic Materials}
  \textbf{\bibinfo{volume}{247}}, \bibinfo{pages}{PII S0304}
  (\bibinfo{year}{2002}{\natexlab{a}}).

\bibitem[{\citenamefont{Sun et~al.}(2001)\citenamefont{Sun, Slonczewski,
  Trouilloud, Abraham, Bacchus, Parkin, and Koch}}]{RefWorks:16}
\bibinfo{author}{\bibfnamefont{J.~Z.} \bibnamefont{Sun}},
  \bibinfo{author}{\bibfnamefont{J.~C.} \bibnamefont{Slonczewski}},
  \bibinfo{author}{\bibfnamefont{P.~L.} \bibnamefont{Trouilloud}},
  \bibinfo{author}{\bibfnamefont{D.}~\bibnamefont{Abraham}},
  \bibinfo{author}{\bibfnamefont{I.}~\bibnamefont{Bacchus}},
  \bibinfo{author}{\bibfnamefont{S.~S.~P.} \bibnamefont{Parkin}},
  \bibnamefont{and} \bibinfo{author}{\bibfnamefont{R.~H.} \bibnamefont{Koch}},
  \bibinfo{journal}{Applied Physics Letters} \textbf{\bibinfo{volume}{78}}
  (\bibinfo{year}{2001}).

\bibitem[{\citenamefont{Sun et~al.}(2002{\natexlab{b}})\citenamefont{Sun, Chen,
  Suzuki, Parkin, and Koch}}]{RefWorks:13}
\bibinfo{author}{\bibfnamefont{J.}~\bibnamefont{Sun}},
  \bibinfo{author}{\bibfnamefont{L.}~\bibnamefont{Chen}},
  \bibinfo{author}{\bibfnamefont{Y.}~\bibnamefont{Suzuki}},
  \bibinfo{author}{\bibfnamefont{S.}~\bibnamefont{Parkin}}, \bibnamefont{and}
  \bibinfo{author}{\bibfnamefont{R.}~\bibnamefont{Koch}},
  \bibinfo{journal}{Journal of Magnetism and Magnetic Materials}
  \textbf{\bibinfo{volume}{247}}, \bibinfo{pages}{L237}
  (\bibinfo{year}{2002}{\natexlab{b}}).

\bibitem[{\citenamefont{Mangin et~al.}(2006)\citenamefont{Mangin, Ravelosona,
  Katine, Carey, Terris, and Fullerton}}]{RefWorks:20}
\bibinfo{author}{\bibfnamefont{S.}~\bibnamefont{Mangin}},
  \bibinfo{author}{\bibfnamefont{D.}~\bibnamefont{Ravelosona}},
  \bibinfo{author}{\bibfnamefont{J.~A.} \bibnamefont{Katine}},
  \bibinfo{author}{\bibfnamefont{M.~J.} \bibnamefont{Carey}},
  \bibinfo{author}{\bibfnamefont{B.~D.} \bibnamefont{Terris}},
  \bibnamefont{and} \bibinfo{author}{\bibfnamefont{E.~E.}
  \bibnamefont{Fullerton}}, \bibinfo{journal}{Nature Materials}
  \textbf{\bibinfo{volume}{5}} (\bibinfo{year}{2006}).

\bibitem[{\citenamefont{Baibich et~al.}(1988)\citenamefont{Baibich, Broto,
  Fert, Vandau, Petroff, Eitenne, Creuzet, Friederich, and
  Chazelas}}]{RefWorks:22}
\bibinfo{author}{\bibfnamefont{M.~N.} \bibnamefont{Baibich}},
  \bibinfo{author}{\bibfnamefont{J.~M.} \bibnamefont{Broto}},
  \bibinfo{author}{\bibfnamefont{A.}~\bibnamefont{Fert}},
  \bibinfo{author}{\bibfnamefont{F.~N.} \bibnamefont{Vandau}},
  \bibinfo{author}{\bibfnamefont{F.}~\bibnamefont{Petroff}},
  \bibinfo{author}{\bibfnamefont{P.}~\bibnamefont{Eitenne}},
  \bibinfo{author}{\bibfnamefont{G.}~\bibnamefont{Creuzet}},
  \bibinfo{author}{\bibfnamefont{A.}~\bibnamefont{Friederich}},
  \bibnamefont{and} \bibinfo{author}{\bibfnamefont{J.}~\bibnamefont{Chazelas}},
  \bibinfo{journal}{Physical Review Letters} \textbf{\bibinfo{volume}{61}}
  (\bibinfo{year}{1988}).

\bibitem[{\citenamefont{Binasch et~al.}(1989)\citenamefont{Binasch, Grunberg,
  Saurenbach, and Zinn}}]{RefWorks:23}
\bibinfo{author}{\bibfnamefont{G.}~\bibnamefont{Binasch}},
  \bibinfo{author}{\bibfnamefont{P.}~\bibnamefont{Grunberg}},
  \bibinfo{author}{\bibfnamefont{F.}~\bibnamefont{Saurenbach}},
  \bibnamefont{and} \bibinfo{author}{\bibfnamefont{W.}~\bibnamefont{Zinn}},
  \bibinfo{journal}{Physical Review B} \textbf{\bibinfo{volume}{39}},
  \bibinfo{pages}{4828} (\bibinfo{year}{1989}).

\bibitem[{\citenamefont{Valet and Fert}(1993)}]{RefWorks:24}
\bibinfo{author}{\bibfnamefont{T.}~\bibnamefont{Valet}} \bibnamefont{and}
  \bibinfo{author}{\bibfnamefont{A.}~\bibnamefont{Fert}},
  \bibinfo{journal}{Physical Review B} \textbf{\bibinfo{volume}{48}}
  (\bibinfo{year}{1993}).

\bibitem[{\citenamefont{Pratt et~al.}(1991)\citenamefont{Pratt, Lee, Slaughter,
  Loloee, Schroeder, and Bass}}]{RefWorks:25}
\bibinfo{author}{\bibfnamefont{W.~P.} \bibnamefont{Pratt}},
  \bibinfo{author}{\bibfnamefont{S.~F.} \bibnamefont{Lee}},
  \bibinfo{author}{\bibfnamefont{J.~M.} \bibnamefont{Slaughter}},
  \bibinfo{author}{\bibfnamefont{R.}~\bibnamefont{Loloee}},
  \bibinfo{author}{\bibfnamefont{P.~A.} \bibnamefont{Schroeder}},
  \bibnamefont{and} \bibinfo{author}{\bibfnamefont{J.}~\bibnamefont{Bass}},
  \bibinfo{journal}{Physical Review Letters} \textbf{\bibinfo{volume}{66}}
  (\bibinfo{year}{1991}).

\bibitem[{\citenamefont{Bedau et~al.}(2010{\natexlab{a}})\citenamefont{Bedau,
  Liu, Sun, Katine, Fullerton, Mangin, and Kent}}]{RefWorks:26}
\bibinfo{author}{\bibfnamefont{D.}~\bibnamefont{Bedau}},
  \bibinfo{author}{\bibfnamefont{H.}~\bibnamefont{Liu}},
  \bibinfo{author}{\bibfnamefont{J.~Z.} \bibnamefont{Sun}},
  \bibinfo{author}{\bibfnamefont{J.~A.} \bibnamefont{Katine}},
  \bibinfo{author}{\bibfnamefont{E.~E.} \bibnamefont{Fullerton}},
  \bibinfo{author}{\bibfnamefont{S.}~\bibnamefont{Mangin}}, \bibnamefont{and}
  \bibinfo{author}{\bibfnamefont{A.~D.} \bibnamefont{Kent}},
  \bibinfo{journal}{Applied Physics Letters} \textbf{\bibinfo{volume}{97}},
  \bibinfo{pages}{262502} (\bibinfo{year}{2010}{\natexlab{a}}).

\bibitem[{\citenamefont{Bedau et~al.}(2010{\natexlab{b}})\citenamefont{Bedau,
  Liu, Bouzaglou, Kent, Sun, Katine, Fullerton, and Mangin}}]{RefWorks:27}
\bibinfo{author}{\bibfnamefont{D.}~\bibnamefont{Bedau}},
  \bibinfo{author}{\bibfnamefont{H.}~\bibnamefont{Liu}},
  \bibinfo{author}{\bibfnamefont{J.~J.} \bibnamefont{Bouzaglou}},
  \bibinfo{author}{\bibfnamefont{A.~D.} \bibnamefont{Kent}},
  \bibinfo{author}{\bibfnamefont{J.~Z.} \bibnamefont{Sun}},
  \bibinfo{author}{\bibfnamefont{J.~A.} \bibnamefont{Katine}},
  \bibinfo{author}{\bibfnamefont{E.~E.} \bibnamefont{Fullerton}},
  \bibnamefont{and} \bibinfo{author}{\bibfnamefont{S.}~\bibnamefont{Mangin}},
  \bibinfo{journal}{Applied Physics Letters} \textbf{\bibinfo{volume}{96}},
  \bibinfo{pages}{022514} (\bibinfo{year}{2010}{\natexlab{b}}).

\end{thebibliography}
\end{document}